\documentclass{ws-ijmpcs}

\begin{document}

\markboth{G. Pelletier \& al.}
{Moderately magnetized relativistic shocks}

%
\catchline{}{}{}{}{}
%

\title{Collisionless Relativistic Shocks:\\
Current driven turbulence and particle acceleration
}

\author{Guy Pelletier}

\address{Institut de Plan\'etologie et d'Astrophysique de Grenoble, Domaine Universitaire\\
F-38041 Grenoble, France\\
Guy.Pelletier@obs.ujf-grenoble.fr}

\author{Martin Lemoine}

\address{Institut d'Astrophysique, 98bis bd Arago\\
F-75014, Paris, France\\
lemoine@iap.fr}

\author{Laurent Gremillet}

\address{CEA, DAM, DIF\\
F-91297 Arpajon, France\\
laurent.gremillet@cea.fr}

\author{Illya Plotnikov}

\address{Institut de Plan\'etologie et d'Astrophysique de Grenoble, Domaine Universitaire\\
F-38041 Grenoble, France\\
Illya.Plotnikov@obs.ujf-grenoble.fr}

\maketitle

\begin{history}
\received{Day Month Year}
\revised{Day Month Year}
\end{history}

\begin{abstract}
The physics of collisionless relativistic shocks with a moderate magnetization is presented. Micro-physics is relevant to explain the most energetic radiative phenomena of Nature, namely that of the termination shock of Gamma Ray Bursts.
A transition towards Fermi process occurs for decreasing magnetization around a critical value which
turns out to be the condition for the scattering to break the mean field inhibition. Scattering is produced by magnetic
micro-turbulence driven by the current carried by returning particles, which had not been considered till now, but turns out to be more intense 
than Weibel's one around the transition. The current is also responsible for a buffer effect on the motion of the incoming flow, on which the threshold for the onset of turbulence depends.

\end{abstract}

\ccode{PACS numbers: 95.30.Qd, 98.70.Rz, 52.35.Tc}

\section{Relativistic shock and Fermi process}	

The development of PIC simulations of relativistic shocks\cite{spit,sir,nish} is providing expected and non-expected results that stimulate
theoretical understanding of the physics, especially in the latter case. The magnetization parameter $\sigma$ together with the shock front 
Lorentz factor $\Gamma_s$ are the essential parameters to scan the properties of those shocks; $\sigma$ is defined as the ratio of the flux
of magnetic energy across the shock over the flux of kinetic energy:
\begin{equation}
\label{ }
\sigma \equiv \frac{B_t^2}{4\pi \Gamma_s^2 \rho_0 c^2} = \frac{B_0^2 \sin^2 \theta_B}{4\pi \rho_0 c^2} \ ,
\end{equation}
where $B_0$ is the ambient magnetic field and $\rho_0$ the ambient mass density.
Ultra-relativistic shocks are now fairly well understood in the two extremes, namely, when $\sigma > 0.1$ and when $\sigma < 10^{-5}$, say. At large $\sigma$, the synchrotron maser instability of a coherent extraordinary mode produces 
the shock structure both in a pair plasma\cite{ha,gh} and in a proton plasma\cite{lyu,sir11,plot13b}. When $\sigma$ is very small, the growth of Weibel instability  allows the reflection of part of the incoming flow, this flux of returning particles being responsible for the instability itself. Agreements between numerical simulations and theoretical works are satisfying. For intermediate magnetizations, PIC simulations by Sironi et al., Ref.~\refcite{sir}, provoked theoretical questions, because the transition towards Fermi process with decreasing magnetization does not fit with the Weibel turbulence scenario. In particular the critical value of the magnetization parameter is surprisingly independent of the shock Lorentz factor.
In that presentation we proposed a consistent theory of the shock structure and the excitation of micro-turbulence that fits 
with the numerical results. The theory focuses on the essential role played by the transverse current carried by the returning particles in shaping the shock forefront, so-called the shock ``foot", and in triggering magnetic turbulence; this will be detailed in a forthcoming paper Ref.~\refcite{lem13a}

\section{Shock ``foot"}

A collisionless shock is built by the growth of an electromagnetic field that reflects part of the incoming particles. Thus a back-stream of returning particles, namely reflected particles and possible up-scattered particles undergoing a Fermi cycle, interact with the incoming flow. The foot length 
is on order of the Larmor radius of the returning particles measured in the front frame. The interaction
occurs through two processes, on one hand, a two-stream interaction that can excite Weibel filamentation after a stage of electrostatic instabilities, 
on other hand, a current driven filamentation. Indeed that latter process stems from the following situation. The mean field is generically almost perpendicular to the shock normal in the front frame of an ultra-relativistic shock. Then returning particles of opposite charge rotate in opposite directions in the mean field and thus generate an electric current perpendicular to the mean field and to the shock normal. That current is compensated by a current carried by the incoming plasma and that compensating current triggers a filamentation instability. These two types of filamentation instabilities have similarities and differences that will be presented in section 4.

Before looking at the excitation of magnetic turbulence, let us examine the flows in the shock foot. We assume that the shock front moves in the x-direction at a velocity $V_s \vec e_x$, the ambient magnetic field is in the z-direction, so that the current carried by returning particles is such that 
$\vec J^{cr} = -\xi_{cr} nec \vec e_y$, where $n = \Gamma_s n_0$ and $\xi_{cr}$ is the fraction of incoming energy density converted into supra-thermal pressure. The compensating current is thus positively oriented along the y-direction. The incoming plasma then suffers a magnetic braking with respect to the shock front, which modifies the global Lorentz factor of the flow and the Lorentz factor of the centre of mass. The global Lorentz factor is slightly modified according to:
\begin{equation}
\label{ }
\Delta \Gamma \sim \frac{JBr_L}{cn_0\Gamma_s m_pc^2} \sim \xi_{cr} \Gamma_s \ .
\end{equation}
Now the motion of the centre of mass is modified differently for a pair plasma and for a proton plasma. In a pair plasma, incoming electrons and positrons are deflected in the y-direction with opposite velocities to produce the compensating current and $v_y \sim \pm \xi_{cr}$. The centre of mass is therefore slowed down\cite{lem13a} such that
\begin{equation}
\label{ }
\Gamma_{cm} \sim \frac{\Gamma_s}{\sqrt{1+\Gamma_s^2 \xi_{cr}^2}} \ .
\end{equation}
Thus, in a mildly relativistic shock such that $\Gamma_s < 1/ \xi_{cr}$, $\Gamma_{cm} \simeq \Gamma_s$, whereas in an ultra-relativistic shock such that $\Gamma_s \gg 1/ \xi_{cr}$, $\Gamma_{cm} \sim 1/\xi_{cr}$.

In a protonic plasma, if the incoming electrons are rapidly heated up to rough equipartition with protons, the behavior is similar to the case of a pair plasma. If the electrons do not reach such a high temperature, because the foot length is too short, then numerical simulation\cite{sir} shows that electrons are not reflected back. Then the situation changes: the foot carries a positive electric charge, say $\rho_{el} \sim \xi_{cr} ne$.
It can be shown\cite{lem13b} that the Lorentz force density in the y-direction, $\rho_{el} E_y - J_xB_z/c \simeq \xi_{cr} \beta_s neB$, deviates protons such that
they acquire a transverse velocity $v_y \sim \xi_{cr}$. Without an electric charge in the foot, the incoming protons could not be significantly deviated.
So again, the centre of mass of the incoming flow is slowed down with respect to the front and the foot plays the role of a buffer which reduces
the Lorentz factor of the ultra relativistic incoming flow to the value $1/\xi_{cr}$ for all $\Gamma_s \gg 1/\xi_{cr}$. This buffer effect will turn out to be crucial in determining the transition towards Fermi process, as will be seen in section 4.

\section{Scattering limit due to the mean field}

Because the transverse component of the mean field is amplified by the Lorentz factor $\Gamma_s$ of the transformation from the ambient (upstream) rest frame to the shock front frame, relativistic shocks are generically superluminal and the mean field, almost perpendicular to the flow direction, drags particles in the downstream flow so that only part of them can come back upstream and only once\cite{beg}. 
In order to get many cycles across the front and have an operative Fermi process, a fast scattering process in required. However even very intense, a usual large scale turbulence cascading towards small scale is unable to make the Fermi process operative in a relativistic shock, because
particles penetrating from downstream to upstream experiences a large scale transverse field during a very short time and over a short length, a Larmor scale as measured in the front frame\cite{nie,lem6}. Fermi process is operative when an intense short scale magnetic field
scatters particles in a time shorter than the Larmor time\cite{pell9}. The scattering frequency in short scale turbulence decreases with the inverse of the particle energy to the square. Precisely the scattering law, proposed in Ref.~\refcite{pell9}, is given by the following equation, as checked by numerical simulations\cite{plot11,sir}:
\begin{equation}
\label{ }
\nu_s = e^2 <\delta B^2> \tau_c/p^2 \ ,
\end{equation}
with $\tau_c \sim \ell_c/c$, $\ell_c$ being the correlation length of the micro-turbulent magnetic field. Since the Larmor frequency decreases inversely proportional to the particle energy, there exists a maximum energy above which a particle can no longer be accelerated by the Fermi process. Fermi process is possible when $\sigma < \xi_B^2$, $\xi_B$ being the fraction of the incoming energy converted into magnetic turbulence, and the maximum extension of the supra-thermal tail (measured in upstream rest frame), due to the scattering limit only, is given by
\begin{equation}
\label{ }
\gamma_{max} \simeq \Gamma_s {\xi_B \over \sqrt{\sigma}} \ .
\end{equation}
For this reason, it is impossible to accelerate protons beyond an energy of $10^{16} eV$ in the termination shock of Gamma Ray Bursts with Fermi process. However electrons are accelerated to very high energies and, as will be seen, can account for the tremendous radiation with the intense
magnetic micro-turbulence. See Ref.~\refcite{pell11,plot13}, confirmed by Ref.~\refcite{sir}.

\section{The two Filamentation Instabilities}

Let us briefly present the two relevant electromagnetic instabilities that generate magnetic turbulence in the precursor of a relativistic shock of low magnetization ($\sigma < 0.1$). For a sake of simplicity, we will compare the two electromagnetic instabilities in a pair plasma\cite{lem13a}. Their development in a protonic plasma and the study of the nonlinear evolution of these instabilities will be presented in a forthcoming paper\cite{lem13b}.

Weibel Filamentation Instability (WFI) is triggered when a tenuous relativistic stream of pairs is pervading an ambient pair plasma. Suppose that a transverse magnetic perturbation has been
generated. It separates pairs of the beam through the force $\pm e \vec v_0 \times  \delta \vec B$, which generates a wavy electric current.
That current, in turn, generates a transverse magnetic perturbation that turns out to be in phase with the original one; therefore the instability.
Its growth rate $g(k)$, for a transverse wave vector of modulus $k$, is maximum for $k^2\delta_e^2 >1$ with $g_{max} = \xi_{cr}^{1/2} \omega_{pe}$ and decreases as $k\delta_e g_{max}$ for smaller wave number.
The nonlinear evolution of the instability creates filaments that are paired with current of opposite direction.

The Current driven Filamentation Instability (CFI) is triggered by the current carried by returning particles. In this case, a transverse magnetic perturbation does not separate charges, but produces a magnetic pressure gradient that perturbs the pair density of the incoming flow. Thus the 
compensating current is perturbed, which generates a perturbed magnetic field, and so amplifies the previous one. The growth rate of
the CFI is similar to the WFI one, but stronger with $g_{max} = \omega_{pe}$.
The nonlinear evolution of the instability creates filaments that are concentrations of the electric current.

\section{Transition towards Fermi process} 

First, we consider the case of a pair plasma. The growth length of turbulence in the foot, $V_{cm}/g_{max}$, has to be shorter than the length of the foot measured in co-moving frame, $c\tau_L/\Gamma_{cm}$, where $\tau_L$ is the Larmor time of the returning particles measured in the front frame; $\tau_L = \Gamma_{cm}mc/e\Gamma_{cm} B_0 = \omega_c^{-1} \equiv mc/eB_0$. The condition for growing Weibel turbulence, i.e. $g_{max} > \Gamma_{cm}V_{cm} \omega_c$ is thus
$\sigma < \xi_{cr} /\Gamma_{cm}^2$ or, in ultra-relativistic regime, $\sigma < \xi_{cr}^3$. The condition with CFI turbulence is less severe with $\sigma < 1/\Gamma_{cm}^2$ or, in ultra-relativistic regime, $\sigma < \xi_{cr}^2$.
Therefore we have a sequence of transitions in a pair plasma for decreasing magnetization\cite{lem13a}: when $\sigma$ becomes smaller than $\xi_{cr}^2$ (typically $10^{-2}$), magnetic micro-turbulence is excited, then for lower $\sigma$, below $\xi_B^2$ (between  $10^{-2}-10^{-3}$, since $\xi_B \lesssim \xi_{cr}$), scattering is strong enough to allow an operative Fermi process, and when $\sigma < \xi_{cr}^3$ (about $10^{-3}$), Weibel turbulence is excited. The buffer effect in the shock foot makes these transitions independent of $\Gamma_s$ in ultra-relativistic regime.
The conclusions are the same in a protonic plasma if the electrons are heated up to rough equipartition with protons. However, PIC simulations\cite{sir} show that this is not the case when the shock foot is too short ($10^{-3} < \sigma <10^{-2}$) and when the electrons remain cold they are not reflected back at the shock front.

In a proton-electron plasma, the intensity of micro-turbulence ($\xi_B \sim 1-10 \%$) is so high that electrons are strongly shaken up by the associated electric field in a relativistic regime. This is characterized by an intensity parameter $a$ defined by
\begin{equation}
\label{ }
a \equiv \frac{eE_{rms}}{m_e \omega_0 c} \ ,
\end{equation}
that is very large. The electron temperature rapidly rises up to a value $T_e = a m_ec^2$. This solves the issue of electrons heating in relativistic shocks, where they can reach a rough equipartition with protons\cite{plot13}. However this is not the case when the magnetization is below but close to its critical value.
Indeed when electrons remain cold, they are not reflected back, so that the foot is filled out with returning protons. That electric charge slows down the incoming plasma such that, again in ultra-relativistic regime, $\Gamma_{cm} \simeq 1/\xi_{cr}$; a buffer effect develops also, as indicated in section 2. A detailed analysis of the excitation of whistler waves will be published in a forthcoming paper Ref.~\refcite{lem13b}.

\section{Performances of the termination shock of GRBs and conclusion}

As already mentioned and published\cite{pell9,pell11,plot13} (see Lemoine presentation in this meeting and references therein), the termination shock of GRBs cannot produce cosmic rays beyond $10^{16} eV$, the expansion and escape losses putting limitation around an energy on the same order of magnitude. Thus  ultra high energy cosmic rays are probably produced during the prompt stage by internal shocks, but not at the termination shock. However the high energy performances of the termination shock have to be emphasized for accounting for the most intense radiation events of Nature, due to both electron energization and magnetic field generation. Indeed the conversion ratio into magnetic micro-turbulence $\xi_B$ is close to $\xi_{cr}$\cite{plot13} and PIC simulations indicate that $\xi_{cr}$ is on the order of 10 percent. Despite its short coherence scale, the magnetic disturbances are generated at a so high level that high energy electrons
radiate a synchrotron-like emission because the so-called wiggler parameter $a_w$ is large:
\begin{equation}
\label{ }
a_w \equiv \frac{eB_{rms} \ell_c}{m_ec^2} \gg 1 \ .
\end{equation}
Note that in a relativistic shock both parameters $a$ and $a_w$ have the same value. Synchrotron diagnostics are thus probing micro-turbulence\cite{lem13}.
The limitation of electron energization due to synchrotron loss leads to a cut off of their energy distribution that is independent of the magnetic field intensity, because the acceleration rate and the loss rate are both proportional to the magnetic energy density. Thus the maximum electron Lorentz factor only depends on the electron density at a power $1/6$ in the shock frame\cite{kirk}, which leads to an almost universal limit of $\gamma_{max} \sim 10^6-10^7$.  Therefore the maximum energy of the emitted photons for the observer depends only on $\xi_B^{1/2}$ and $\Gamma_s$.
For $\Gamma_s \sim 300$, that maximum photon energy is of a few GeV\cite{pell11,plot13}. Beyond that energy, the emission is compatible with an SSC process\cite{wang}.

Therefore the most energetic radiation events in the Universe due to the termination shock of GRBs can be entirely explained by plasma micro-physics.


\end{document}